\documentclass{svmult}

\usepackage{mathptmx}       
\usepackage{helvet}         
\usepackage{courier}        
\usepackage{type1cm}        
\usepackage{makeidx}         
\usepackage{graphicx}        
\usepackage{multicol}        
\usepackage[bottom]{footmisc}

\makeindex             

\newcommand{\rhop}{\rho_{\rm p}}
\newcommand{\rhoc}{\rho_{\rm c}}
\newcommand{\etac}{\eta_{\rm C}}
\newcommand{\taud}{\tau_{\rm D}}
\newcommand{\va}{v_{\rm A}}
\newcommand{\rhoe}{\rho_{\rm e}}
\newcommand{\lp}{L_{\rm p}}

\begin{document}

\title*{Magnetohydrodynamic Waves in Partially Ionized Prominence Plasmas}
\titlerunning{MHD waves in partially ionized prominence plasmas} 
\author{R. Soler and J. L. Ballester}
\institute{R. Soler \at Centre for Plasma Astrophysics,Department of Mathematics, KULeuven, \email{roberto.soler@wis.kuleuven.be}
\and J. L. Ballester \at Departament de F\'\i sica, Universitat de les Illes Balears \email{joseluis.ballester@uib.es}}
\maketitle

\abstract{Prominences or filaments are cool clouds of partially ionized plasma living in the solar corona. Ground- and space-based
observations have confirmed the presence of oscillatory motions in
prominences and they have been interpreted in terms of
magnetohydrodynamic (MHD) waves. Existing observational evidence points out that these oscillatory motions are damped in short spatial and temporal scales by some still not well known physical mechanism(s).  Since prominences are partially ionized plasmas, a potential mechanism able to damp these oscillations could be ion-neutral collisions.  Here, we will review the work done on the effects of partial ionization on MHD waves in prominence plasmas.}

\section{Introduction}
\label{sec:intro}
Quiescent solar filaments are clouds of cool and dense plasma suspended against gravity by forces thought to be of magnetic origin. High-resolution H$_\alpha$ observations  (\cite{lin05}, \cite{heinzel06})  have revealed that  the fine structure of filaments is apparently composed by many horizontal and thin dark threads (see \cite{lin11}, for a review).  The measured average width of resolved thin threads is about $0.3$ arc.sec ($\sim$ $210$  km) while their length is between $5$ and $40$ arc.sec ($\sim$ 3500 - 28000  km). The fine threads of solar filaments seem to be partially filled with cold plasma \cite{lin05}, typically two orders of magnitude denser and cooler than the surrounding corona, and it is generally assumed that they outline their magnetic flux tubes \cite{engvold98,lin04, lin05,engvold08,martin08,lin08}. This idea is strongly supported by observations which suggest that they are inclined with respect to the filament long axis in a similar way to what has been found for the magnetic field (\cite{leroy80,bommier94,bommier98}).  

Small amplitude oscillations in prominences and filaments are a commonly observed phenomenon. The detected peak velocity ranges from the noise level (down to 0.1 km~s$^{-1}$ in some cases) to 2--3~km~s$^{-1}$. The observed periodic signals are mainly detected from Doppler velocity measurements and can therefore be associated to the transverse displacement of the fine structures \cite{lin09}.  Two-dimensional observations of filaments \cite{yi91a,yi91b} revealed that individual threads or groups of threads may oscillate independently with their own periods, which range between 3 and 20 minutes.  Furthermore, \cite{Lin07} have shown evidence about traveling waves along a number of filament threads with an average phase velocity of $12$  km s$^{-1}$, a wavelength of $4''$ ($\sim 2800$ \ km), and oscillatory periods of the individual threads that vary from $3$ to $9$ minutes. 
 
Observational evidence for the damping of small amplitude oscillations in prominences can be found in \cite{arregui11a}. Observational studies have allowed to obtain some characteristic spatial and time scales. Reliable values for the damping time have been derived, from different Doppler velocity time series by \cite{Terradas02}, in prominences, and by \cite{lin04} in filaments.  The values thus obtained are usually between 1 and 4 times the corresponding period, and large regions of prominences/filaments display similar damping times. 

Finally, small amplitude oscillations in quiescent filaments have been interpreted in terms of magnetohydrodynamic (MHD) waves \cite{oliver02} and, in many cases, theoretical works studying the damping of prominence oscillations have studied first the effect of a given damping mechanism on MHD waves in a simple, uniform, and unbounded media before to introduce structuring and non-uniformity. This is the approach that we will follow in this paper.

\section{MHD waves in unbounded partially ionized prominence plasmas}
\label{sec:unbound}

Since the temperature of prominences is typically of the order of 10$^4$ K, the prominence plasma is only partially ionized. The exact ionization degree of prominences is unknown and the reported ratio of electron density to neutral hydrogen density \cite{patsourakos02} covers about two orders of magnitude (0.1\,--\,10). Partial ionization brings the presence of neutrals in addition to electrons and ions, thus collisions between the different species are possible. Because of the occurrence of collisions between electrons with neutral atoms and ions, and more importantly between ions and neutrals, Joule dissipation is enhanced when compared with the fully ionized case. A partially ionized plasma can be represented as a single-fluid in the strong coupling approximation, which is valid when the ion density in the plasma is low and the collision time between neutrals and ions is short compared with other time-scales of the problem. Using this approximation it is possible to describe the very low frequency and large-scale fluid-like behaviour of plasmas \cite {goossens03}.

Partial ionization affects the induction equation, which contains additional terms due to the presence of neutrals and a non-zero resistivity \cite{forteza07}. These additional terms account for the processes of ohmic diffusion,  ambipolar diffusion, and Hall's magnetic diffusion. Ohmic diffusion is mainly due to electron-ion collisions and produces magnetic diffusion parallel to the magnetic field lines; ambipolar diffusion is mostly caused by ion-neutral collisions and Hall's effect is enhanced by ion-neutral collisions since they tend to decouple ions from the magnetic field while electrons remain able to drift with the magnetic field \cite{pandey08}. Due to the presence of neutrals, perpendicular magnetic diffusion is much more efficient than longitudinal magnetic diffusion in a partially ionized plasma.  It is important to note that this is so even for a small relative density of neutrals.

\subsection{Homogeneous and unbounded prominence medium}

Several studies have considered the damping of MHD waves in partially ionized plasmas of the solar atmosphere \cite{depontieu01, james03, khodachenko04, leake05}. In the context of solar prominences, \cite{forteza07} derived the full set of MHD equations for a partially ionized, one-fluid hydrogen plasma and applied them to the study of the time damping of linear, adiabatic fast and slow magnetoacoustic waves in an unbounded prominence medium. This study was later extended to the non-adiabatic case, including thermal conduction by neutrals and electrons and radiative losses \cite{forteza08}. \cite{forteza07} considered a uniform and unbounded prominence plasma and found that ion-neutral collisions are more important for fast waves, for which the ratio of the damping time to the period is in the range 1 to 10$^5$, than for slow waves, for which values between 10$^4$ and 10$^8$ are obtained. Fast waves are efficiently damped for moderate values of the ionization fraction, while in a nearly fully ionized plasma, the small amount of neutrals is insufficient to damp the perturbations.

In the above studies, a hydrogen plasma was considered, although 90\% of the prominence chemical composition is hydrogen while the remaining 10\% is helium. The effect of including helium in the model of \cite{forteza08} was assessed by \cite{soler10}. The species present in the medium are electrons, protons, neutral hydrogen, neutral helium (He\,{\sc i}) and singly ionized helium (He\,{\sc ii}), while the presence of He\,{\sc iii} is neglected \cite{gouttebroze09}. 

\begin{figure}[htbp]
  \includegraphics[width=0.5\textwidth]{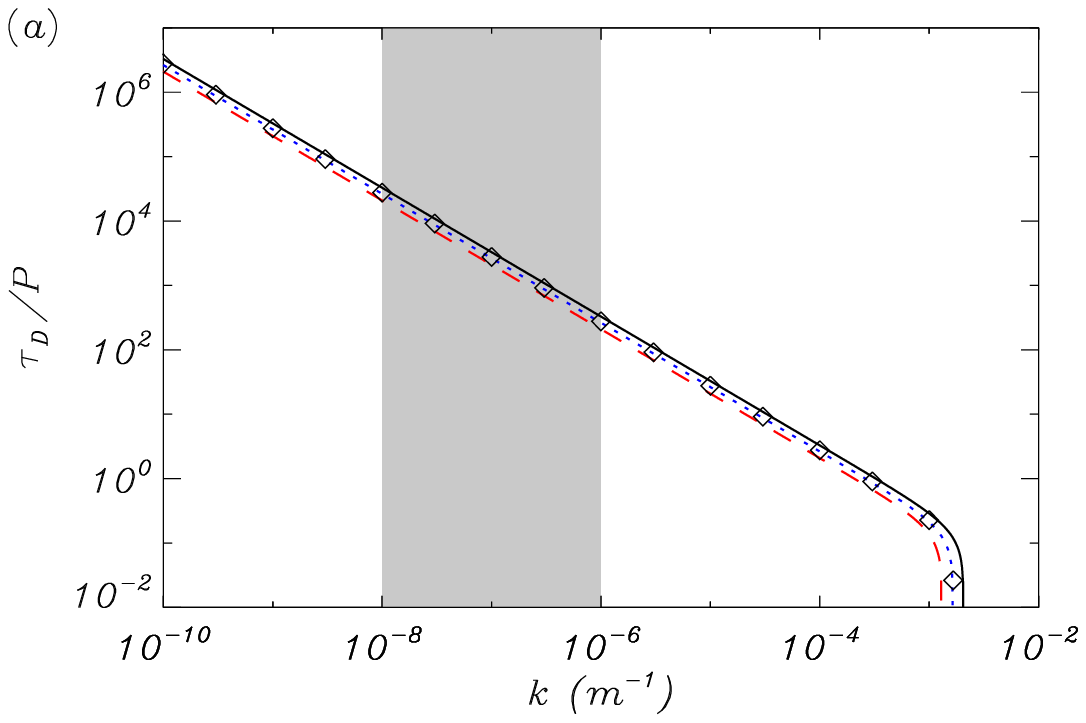}
  \includegraphics[width=0.5\textwidth]{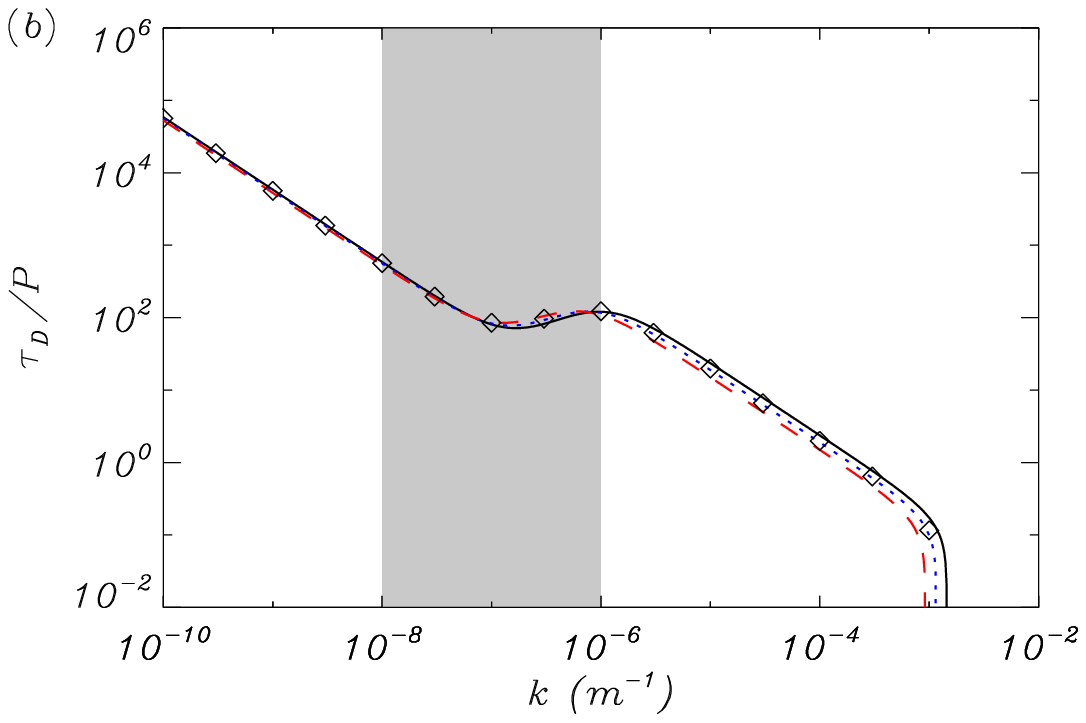}\\
  \includegraphics[width=0.5\textwidth]{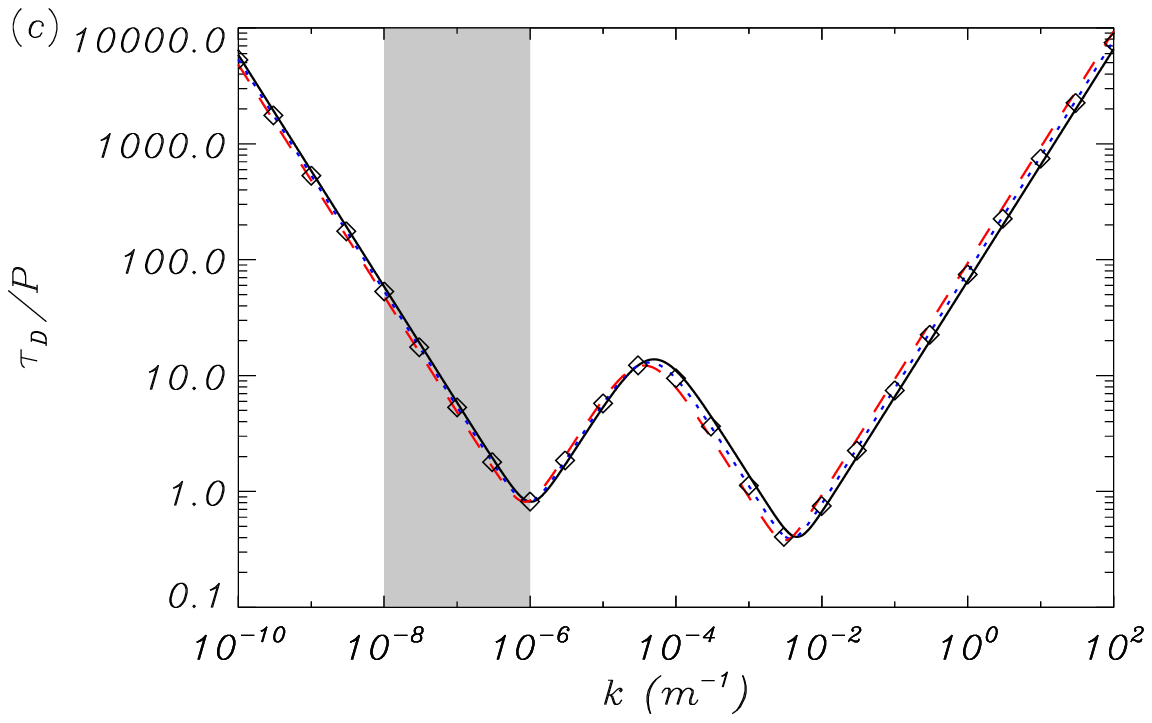}
  \includegraphics[width=0.5\textwidth]{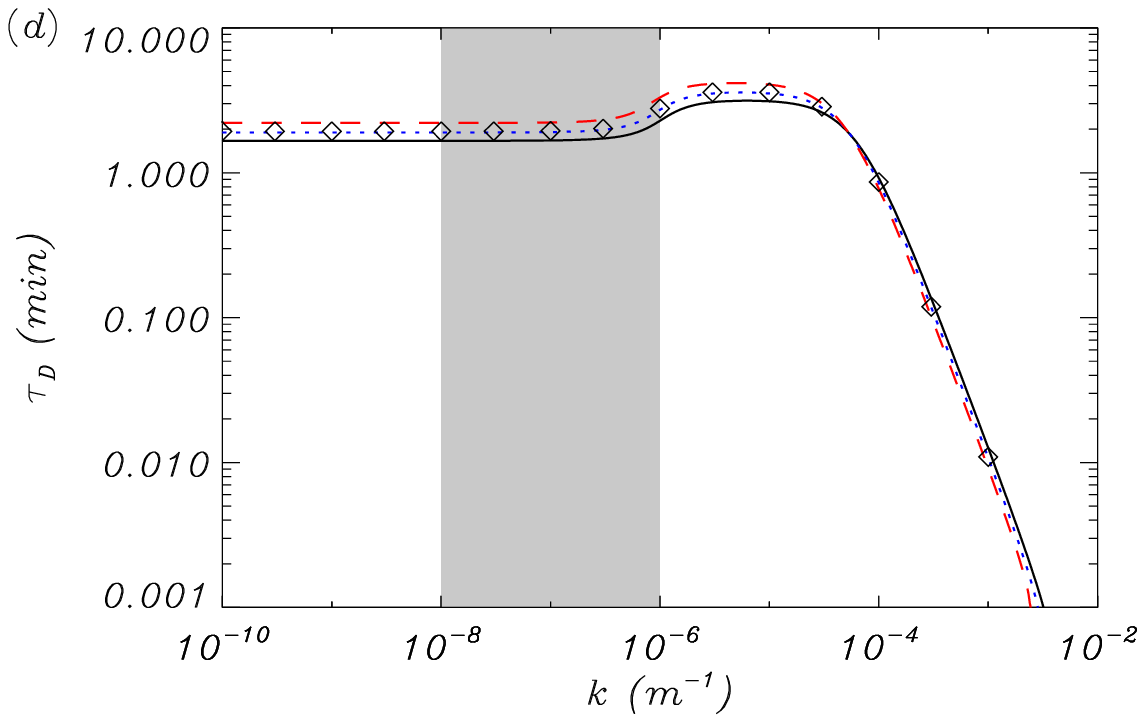}
\caption{Wave damping by ion-neutral effects in a uniform medium. (a)--(c) Ratio of the damping time to the period, $\tau_D/P$, versus the wavenumber, $k$, corresponding to the Alfv\'en wave, fast wave and slow wave, respectively. 
(d) Damping time, $\tau_D$, of the thermal wave versus the wavenumber, $k$. The different linestyles represent the following abundances: $\xi_{\mathrm{HeI}} = 0\%$ (solid line), $\xi_{\mathrm{HeII}} = 10\%$ (dotted line) and $\xi_{\mathrm{HeI}} = 20\%$ (dashed line). In all computations, $\tilde \mu_{\mathrm{H}} = 0.8$ and $\delta_{\mathrm{He}} = 0.1$. The results for $\xi_{\mathrm{HeI}} = 10\%$ and $\delta_{\mathrm{He}} = 0.5$ are plotted by means of symbols for comparison. The shaded regions correspond to the range of typically observed wavelengths of prominence oscillations. In all the figures shown, the angle, $\theta$, between the wavevector and the magnetic field is $\pi/4$. From \cite{soler10}}
\label{fig:mhdwaves}
\end{figure}

The hydrogen ionization degree is characterized by $\tilde \mu_{\mathrm{H}}$ which is equivalent to the mean atomic weight of a pure hydrogen plasma and ranges between 0.5 for fully ionized hydrogen and 1 for fully neutral hydrogen. The helium ionization degree is characterized by $\delta_{\mathrm{He}} = \frac{\xi_{\mathrm{HeII}}}{\xi_{\mathrm{HeI}}}$, where $\xi_{\mathrm{HeII}}$ and $\xi_{\mathrm{HeI}}$ denote the relative densities of single ionized and neutral helium, respectively. Figure~\ref{fig:mhdwaves} a, b, c displays $\tau_D/P$ as a function of the wavenumber, $k$, for the Alfv\'en, fast and slow waves. In this Figure, the results corresponding to several helium abundances are compared for hydrogen and helium ionization degrees of $\tilde \mu_{\mathrm{H}} = 0.8$ and $\delta_{\mathrm{He}}=0.1$, respectively, and it can be observed that the presence of helium has a minor effect on the results. 

The thermal mode is a purely damped, non-pr\-o\-pa\-ga\-ting disturbance ($\omega_r = 0$), so only the damping time, $\tau_D$, is plotted (Figure~\ref{fig:mhdwaves}d). We observe that the effect of helium is different in two ranges of $k$. For $k > 10^{-4}$~m$^{-1}$, thermal conduction is the dominant damping mechanism, so the larger the amount of helium, the shorter $\tau_D$ because of the enhanced thermal conduction by neutral helium atoms. On the other hand, radiative losses are more relevant for $k < 10^{-4}$~m$^{-1}$. In this region, the thermal mode damping time grows as the helium abundance increases.  Since these variations in the damping time are very small, we again conclude that the damping time obtained in the absence of helium does not significantly change when helium is taken into account. Therefore, the inclusion of neutral or single ionized helium in partially ionized prominence plasmas does not modify the behaviour of linear, adiabatic or non-adiabatic MHD waves already found by \cite{forteza07} and \cite{forteza08}. On the other hand, in Figure~\ref{fig:mhdwaves}c we can observe that in the case of slow waves, and within most of the interval of observed wavelengths in prominence oscillations, the ratio between the damping time and the period agrees with the observational determinations, which is due to the joint effect of ion-neutral collisions and non-adiabatic effects (\cite{forteza08}, \cite{soler10}).

\section{Magnetohydrodynamic Waves in Prominence Threads}
\label{sec:threads}

In this Section, we summarize the results of papers which investigate the damping of MHD waves in partially ionized prominence thread models . For simplicity, early investigations neglected the variation of density along the thread and took into account the variation of density in the transverse direction only. Subsequent works incorporated longitudinal inhomogeneity in addition to transverse inhomogeneity.

\subsection{Longitudinally Homogeneous Thread Models}
\label{sec:threadshom}

\begin{figure}[!tb]
\centering
\includegraphics[scale=.4]{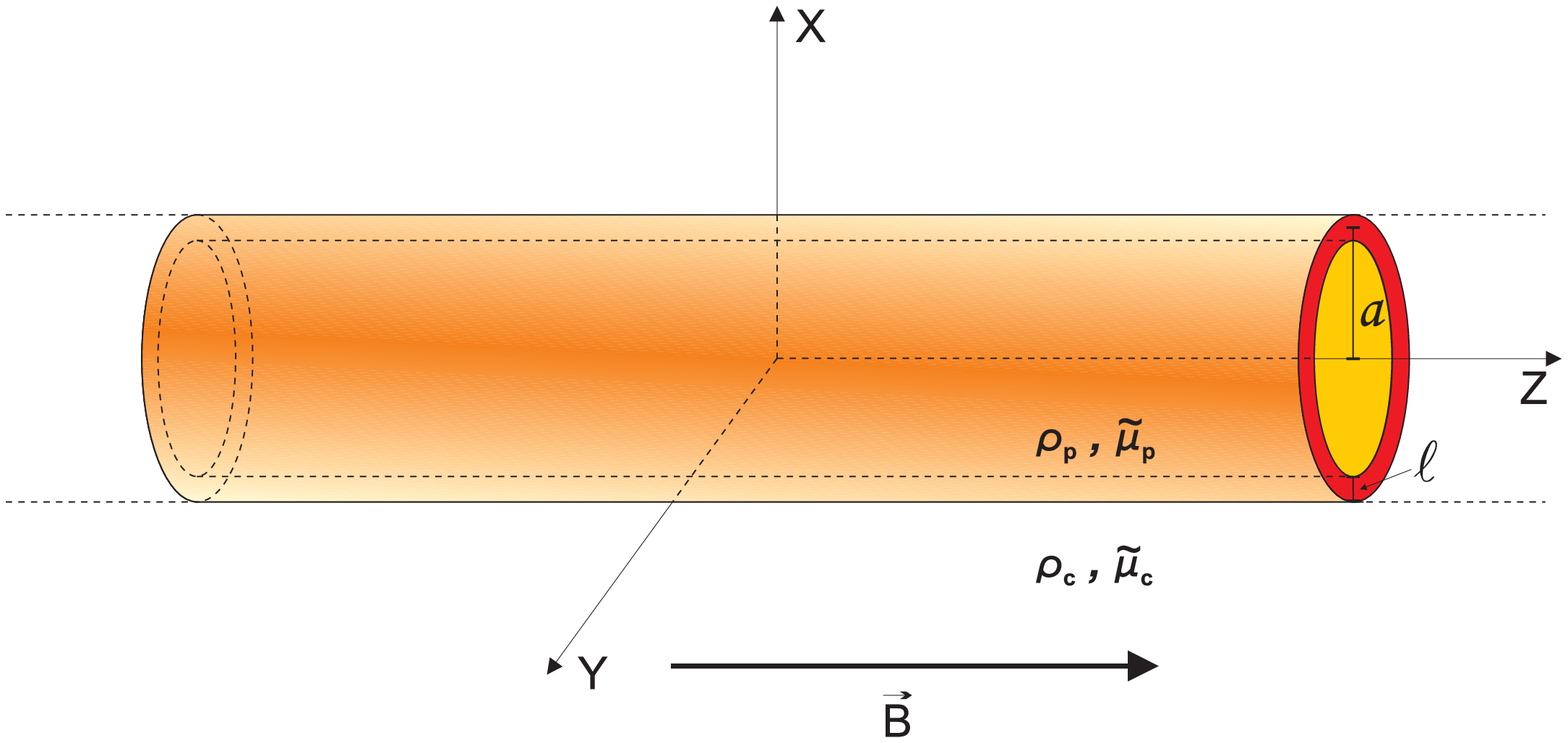}
%
%
\caption{Sketch of the longitudinally homogeneous prominence thread model used in \cite{soler09c} with $l=0$ and in \cite{soler09e,soler11} with $l\neq0$. In this Figure $a$ denotes the radius of the cylinder while in the text we use $R$. Adapted from \cite{soler11}.}
\label{fig:modelhomogeni}       
\end{figure}

The first papers that  studied partial ionization effects on wave propagation in a longitudinally homogeneous prominence thread model were \cite{soler09c,soler09e,soler11}. These authors investigated linear MHD waves superimposed on a straight magnetic cylinder of radius $R$, representing the thread itself, and embedded in a fully ionized and uniform coronal plasma. Gravity was neglected and the magnetic field was taken constant along the axis of the cylinder.  \cite{soler09c} considered an abrupt jump of density in the transverse direction from the internal (prominence), $\rhop$, to the external (coronal), $\rhoc$, densities at the thread boundary, while \cite{soler09e,soler11} replaced the discontinuity in density by a continuous variation of density in a region of thickness $l$. For $l=0$ the equilibrium of \cite{soler09e,soler11} reverts to that of \cite{soler09c}. Hence the ratio $l/R$ indicates the inhomogeneity length-scale in the transverse direction. In both papers the prominence plasma was assumed partially ionized with an arbitrary ionization degree, while the external coronal medium was fully ionized. The single-fluid approximation was adopted and Ohm's, Hall's, and Cowling's terms were included in the induction equation.  Thus, the equilibrium configuration is similar to the classical straight flux tube model investigated by, e.g., \cite{edwin83,goossens09}, with the addition of partial ionization. A sketch of the model is displayed in Figure~\ref{fig:modelhomogeni}.

In this model the observed transverse oscillations of prominence threads can be interpreted in terms of transverse (Alfv\'enic) kink modes. Because of their observational relevance, here we discuss the results for transverse (Alfv\'enic) kink waves only. The interested reader is refereed to the original papers \cite{soler09c,soler09e,soler11} where the results of other waves are explained in detail. It is well known that for $l\neq0$ the kink mode is resonantly coupled to Alfv\'en continuum modes in the region of transversely non-uniform density. As a consequence the kink mode is damped by resonant absorption. In addition, the kink mode is also damped by magnetic diffusion effects due to partial ionization. In the fully ionized case, the ideal resonant damping of the kink mode in prominence threads was investigated by \cite{arregui08,soler09f}, while partial ionization does not affect the mechanism of resonant absorption, which is an ideal process independent of dissipation by ion-neutral collisions. This has been shown by \cite{soler12} using multifluid theory.

\cite{soler09c,soler09e} studied temporal damping of standing waves. By neglecting the effects of Ohm's and Hall's diffusion in comparison to that of Cowling's diffusion, approximate expressions for the period, $P$, and for the ratio of the damping time, $\taud$, to the period of the kink mode can be obtained in the long-wavelength limit, i.e., $\lambda/R \gg 1$, where $\lambda$ is the wavelength. The long-wavelength limit is a reasonable approximation since wavelengths typically observed in prominences are roughly between $10^3$~km and $10^5$~km  (see \cite{oliver02}) while the  observed widths of the threads are between 100~km and 600~km (see \cite{lin11}). The expressions for $P$ and $\taud/P$ are
\begin{eqnarray}
P &=& \frac{\lambda}{\va} \sqrt{\frac{\zeta+1}{2\zeta}}, \label{eq:appper} \\ 
\frac{\taud}{P} &=& \frac{2}{\pi} \left( \frac{l}{R}\frac{\zeta-1}{\zeta+1} +\frac{2 \xi_{\rm n}^2}{1-\xi_{\rm n}} \frac{\omega_{\rm k}}{\nu_{\rm in}}\right)^{-1}, \label{eq:apptdp}
\end{eqnarray}
where $\va$ is the prominence Alfv\'en velocity, $\zeta = \rhop/\rhoc$ is the density contrast, $\omega_{\rm k} = 2\pi/P$ is the kink mode frequency (with $P$ given by Equation~(\ref{eq:appper})),  $\xi_{\rm n}$ is the fraction of neutrals, and $\nu_{\rm in}$ the ion-neutral collision frequency. $\xi_{\rm n} = 0$ for a fully ionized plasma and  $\xi_{\rm n} = 1$ for a neutral medium. To perform a check, we take $\lambda = 10^4$~km, $\va = 50$~km~$^{-1}$, and $\zeta = 200$. Equation~(\ref{eq:appper}) gives $P \approx 2.4$~min, which is consistent with the observed periods. 

Regarding damping, the first term within the parenthesis of Equation~(\ref{eq:apptdp}) is due to resonant absorption and the second term is due to Cowling's diffusion.  Note that the original expression of $\taud/P$ given in \cite{soler09e} involves  Cowling's diffusivity, $\etac$. In the present discussion we have replaced $\etac$ by its expression in terms of $\xi_{\rm n}$ and $\nu_{\rm in}$ (see the expression of $\etac$ in, e.g., \cite{soler09c}). Our purpose is to show that the term related to Cowling's diffusion is proportional to the ratio $\omega_{\rm k}/\nu_{\rm in}$. To perform a simple order-of-magnitude estimation of the importance of Cowling's diffusion, let us take a period of 3~min and compute $\nu_{\rm in}$  using $\xi_{\rm n} = 0.5$, a prominence density of $5\times10^{-11}$~kg~m$^{-3}$, and a prominence temperature of $8,000$~K (see the expression of $\nu_{\rm in}$ in \cite{soler09c}). The result is $\omega_{\rm k}/\nu_{\rm in} \approx 2.38\times10^{-4}$. This estimation indicates that the effect of Cowling's diffusion is negligible unless the prominence is almost neutral, i.e., $\xi_{\rm n} \approx 1$, which is an unrealistic limit. Therefore, resonant absorption dominates the kink mode damping  and the second term within the parenthesis of Equation~(\ref{eq:apptdp}) can be dropped. Hence, for $\zeta = 200$ and $l/R=0.2$ we obtain $\taud/P \approx 3.22$, which again is consistent with the observations.

\begin{figure}[!tb]
\includegraphics[width=.47\columnwidth]{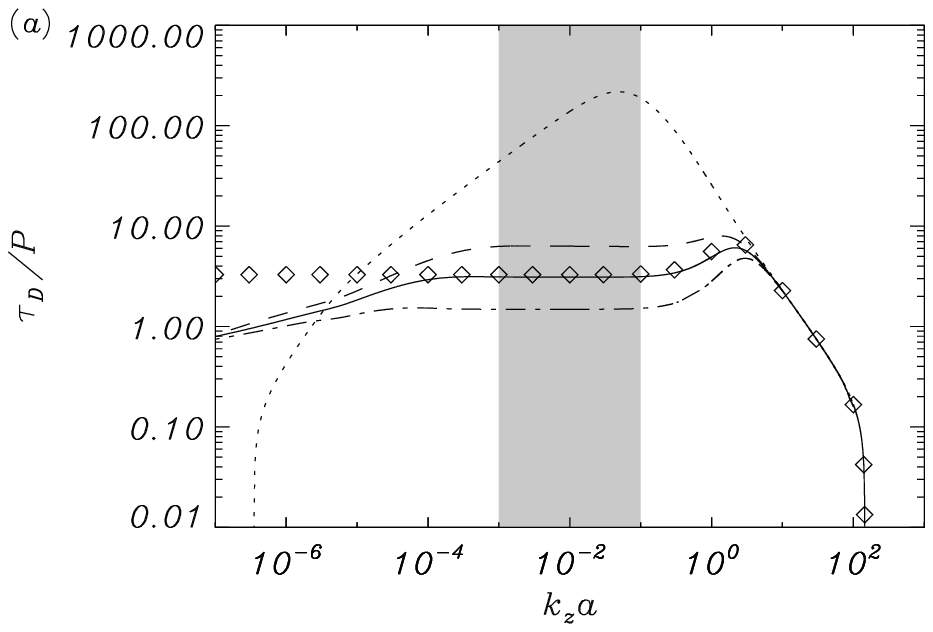}
\includegraphics[width=.45\columnwidth]{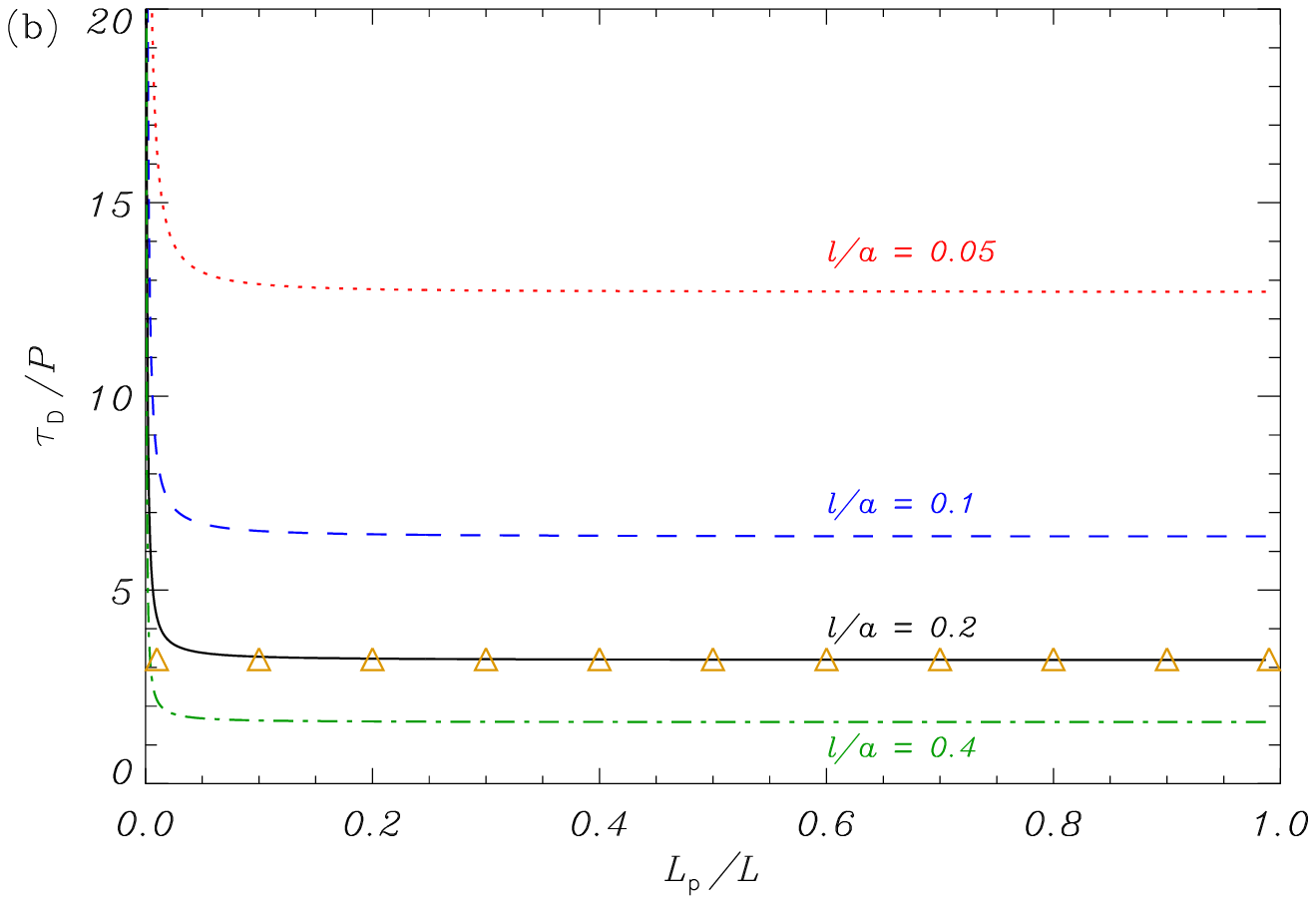}
%
%
\caption{(a) $\taud/P$ vs. $k_za$ for the kink mode in a longitudinally homogeneous thread with $l/a = 0$ (dotted), $l/a = 0.1$ (dashed), $l/a = 0.2$ (solid), and $l/a=0.4$ (dash-dotted). The shaded zone denotes realistic wavelengths. Adapted from \cite{soler09e}. (b)  $\taud/P$ vs. $\lp/L$ for the kink mode in a longitudinally inhomogeneous thread with $l/a = 0.05$ (dotted), $l/a = 0.1$ (dashed), $l/a = 0.2$ (solid), and $l/a=0.4$ (dash-dotted). Symbols are the result from Equation~(\ref{eq:apptdp}) with $l/a = 0.2$. In this Figure $a$ denotes the radius of the cylinder while in the text we use $R$. Adapted from \cite{soler10b}.}
\label{fig:kinkthread}       
\end{figure}

\cite{soler09e} also obtained results beyond the long-wavelength approximation by means of full numerical solutions. They included Ohm's and Hall's terms along with Cowling's diffusion. \cite{soler09e} computed the kink mode $\taud/P$ as a function of the parameter $k_z R$, where $k_z = 2\pi/\lambda$ (see Fig.~\ref{fig:kinkthread}a).  They concluded that Hall's term is always negligible in prominence conditions, Ohm's diffusion is only important for extremely long wavelengths (very small values of $k_z R$), and Cowling's diffusion is only relevant for short wavelengths (large $k_z R$). For realistic wavelengths, i.e., $10^{-3} < k_zR < 10^{-1}$, resonant absorption determines the damping rate of the kink mode and the analytical formula given in  Equation~(\ref{eq:apptdp}) is very accurate. 

Subsequently, \cite{soler11} used the same model as \cite{soler09e} to study spatial damping of kink waves. The results of \cite{soler11} are qualitatively equivalent to those of \cite{soler09e}, i.e., resonant damping dominates for realistic frequencies whereas  Cowling's diffusion is efficient for high frequencies only.

\subsection{Longitudinally Inhomogeneous Thread Models}
\label{sec:threadsin}

A longitudinally homogeneous cylinder is a crude representation of a prominence fine structure. High-resolution observations of prominences (see \cite{lin11}) suggest that the cool and dense material of the threads only occupies a small part of longer magnetic flux tubes, with the rest of the magnetic tube probably filled with hot coronal plasma. The observed length of the threads is believed to be only a small percentage of the total length of the magnetic tube, whose feet are rooted in the solar photosphere \cite{ballester89,rempel99}. This observational evidence may be omitted to study propagating waves in the dense part of the magnetic tube if the wavelengths are much shorter than the length of the threads. For this case the longitudinally homogeneous models discussed in Section~\ref{sec:threadshom} may be appropriate. However, in the case of standing modes, the associated wavelengths are of the order of the total length of magnetic field lines. Therefore the longitudinal structuring of the prominence magnetic tube cannot be neglected when standing modes are investigated.

\begin{figure}[!tb]
\centering
\includegraphics[scale=.4]{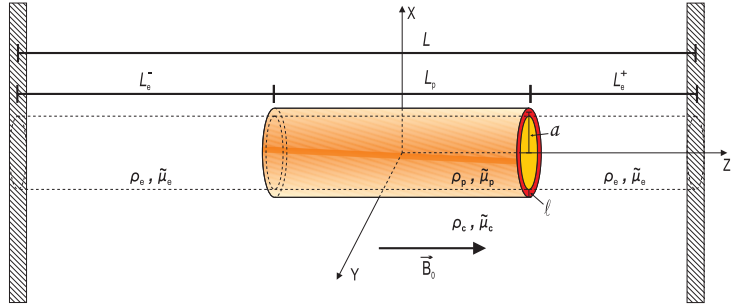}
%
%
\caption{Sketch of the longitudinally inhomogeneous prominence thread model used in \cite{soler10b}. In this Figure $a$ denotes the radius of the cylinder while in the text we use $R$. Adapted from \cite{soler10b}.}
\label{fig:modelinhomogeni}       
\end{figure}

 This observational evidence has been taken into account in some works  which studied ideal, undamped kink modes in the fully ionized case (see, e.g., \cite{diaz02,terradas08,soler11b}). The first paper that incorporated the effects of damping by Cowling's diffusion and resonant absorption on standing kink modes in longitudinally inhomogeneous threads was \cite{soler10b}.  These authors used the model displayed in Figure~\ref{fig:modelinhomogeni}. It is composed of a straight magnetic cylinder of length $L$ and radius $R$ whose ends are fixed at two rigid walls representing the solar photosphere. The magnetic field is constant. The cylinder is composed of a region of length $\lp$ and density $\rhop$, representing the prominence thread, surrounded by two regions of density $\rhoe$ representing the evacuated part of the tube. The external coronal density is $\rhoc$ and for simplicity it is set $\rhoe = \rhoc$. The prominence thread is transversely inhomogeneous in a region of thickness $l$ where the density continuously varies from $\rhop$ to $\rhoc$. The prominence plasma is partially ionized while the coronal and evacuated plasmas are fully ionized.

To investigate standing kink modes analytically, \cite{soler10b} used the thin tube approximation, i.e., $R/L \ll 1$ and $R/\lp \ll 1$. To check this approximation we take the values of $R$ and $\lp$ typically reported from the observations (see \cite{lin11}) and assume $L \sim 10^5$~km, so that $R/\lp$ and $R/L$ are in the ranges $2\times 10^{-3} < R/\lp < 0.1$ and $5\times 10^{-4} < R/L < 3\times 10^{-3}$, meaning that the use of the TT approximation is justified. \cite{arregui11c} have shown that the results of \cite{soler10b} remain valid beyond the thin tube approximation. \cite{soler10b} derived approximate expressions for $P$ and the ratio $\taud/P$. The expression for $P$ is
\begin{equation}
 P = \frac{\pi}{\va}  \sqrt{\frac{\zeta+1}{2\zeta}} \sqrt{\left( L-\lp \right) \lp}, \label{eq:inhomoper}
\end{equation}
where here $\va$ is the Alfv\'en velocity of the dense, prominence plasma only, and $\zeta = \rhop / \rhoc$ is the density constrast as before. In Equation~(\ref{eq:inhomoper}) it is assumed that the prominence thread is located at the center of the magnetic tube (a general expression is given in \cite{soler10b}). A direct comparison of Equations~(\ref{eq:appper}) and (\ref{eq:inhomoper}) shows that the effect of the longitudinal structuring of the tube is to select a particular value of the wavelength, $\lambda$, which depends of the relation between $L$ and $\lp$. Regarding the damping rate, the expression for $\taud/P$ is not explicitly given here because it is exactly the same as that in Equation~(\ref{eq:apptdp}), where now $\omega_{\rm k} = 2\pi / P$ has to be computed using the period from Equation~(\ref{eq:inhomoper}). Thus, as happens for kink modes in longitudinally homogeneous threads, the effect of Cowling's diffusion is negligible for realistic values of the period when compared to that of resonant absorption.


Figure~\ref{fig:kinkthread}b shows the ratio $\taud/P$ numerically computed by \cite{soler10b} as a function of $\lp/L$. Remarkably, the damping ratio is independent of the length of the thread and only depends on the transverse non-uniformity length scale $l/R$. These means that the expression of $\taud/P$ for longitudinally homogeneous (Equation~(\ref{eq:apptdp})) tubes also applies when longitudinal structuring is included.

\begin{acknowledgement}
RS acknowledges support from a Marie Curie Intra-European Fellowship within the European Commission 7th Framework Program (PIEF-GA-2010-274716). RS and JLB acknowledge financial support from MICINN and FEDER funds through grant AYA2011-22486.

\end{acknowledgement}

 \bibliographystyle{spphys}
 \bibliography{refs}

\end{document}